\documentclass[amsmath,amssymb,aps,prc,reprint,superscriptaddress]{revtex4-1}
\usepackage[utf8]{inputenc}
\usepackage{booktabs}
\usepackage{graphicx}
\usepackage{subfigure}
\usepackage{float}
\usepackage{xcolor}
\usepackage{dcolumn}
\makeatletter
\newcommand{\nuc}[2]{\ensuremath{^{\text{#1}}\text{#2}}}
\makeatother

\begin{document}
\preprint{v3 - November 1, 2023}
\title{$^{11}$B states above the $\alpha$-decay threshold studied via $^{10}$B$(d,p){}^{11}$B}



\author{A.~N.~Kuchera}
\email[]{Corresponding author: ankuchera@davidson.edu}
\affiliation{Department of Physics, Davidson College, Davidson, North Carolina 28035, USA}

\author{G.~Ryan}
\affiliation{Department of Physics, Davidson College, Davidson, North Carolina 28035, USA}

\author{G.~Selby}
\affiliation{Department of Physics, Davidson College, Davidson, North Carolina 28035, USA}

\author{D.~Snider}
\affiliation{Department of Physics, Davidson College, Davidson, North Carolina 28035, USA}

\author{S.~Anderson}
\email[]{Deceased}
\affiliation{Department of Physics, Davidson College, Davidson, North Carolina 28035, USA}

\author{S. Almaraz-Calderon}
\affiliation{Department of Physics, Florida State University, Tallahassee, Florida 32306, USA}

\author{L.~T.~Baby}
\affiliation{Department of Physics, Florida State University, Tallahassee, Florida 32306, USA}

\author{B.~A.~Brown}
\affiliation{Facility for Rare Isotope Beams, Michigan State University, East Lansing, Michigan 48824, USA}
\affiliation{Joint Institute for Nuclear Astrophysics: Center for the Evolution of the Elements, Michigan State University, East Lansing, Michigan 48824, USA}
\affiliation{Department of Physics and Astronomy, Michigan State University, East Lansing, Michigan 48824, USA}

\author{K.~Hanselman}
\affiliation{Department of Physics, Florida State University, Tallahassee, Florida 32306, USA}

\author{E. Lopez-Saavedra}
\affiliation{Department of Physics, Florida State University, Tallahassee, Florida 32306, USA}

\author{K.~T.~Macon}
\affiliation{Department of Physics and Astronomy, Louisiana State University, Baton Rouge, Louisiana 70803, USA}

\author{G.~W.~McCann}
\affiliation{Department of Physics, Florida State University, Tallahassee, Florida 32306, USA}

\author{K.~W.~Kemper}
\affiliation{Department of Physics, Florida State University, Tallahassee, Florida 32306, USA}

\author{M.~Spieker}
\affiliation{Department of Physics, Florida State University, Tallahassee, Florida 32306, USA}

\author{I.~Wiedenh\"{o}ver}
\affiliation{Department of Physics, Florida State University, Tallahassee, Florida 32306, USA}

\begin{abstract}

The resonance region of $^{11}$B covering excitation energies from 8.4\,MeV to 13.6\,MeV was investigated with the $(d,p)$ reaction performed on an enriched $^{10}$B target at the Florida State University Super-Enge Split-Pole Spectrograph of the John D. Fox Superconducting Linear Accelerator Laboratory. Complementary measurements were performed with a target enriched in $^{11}$B to identify possible $^{12}$B contaminants in the $(d,p)$ reaction. Four strongly populated $^{11}$B states were observed above the $\alpha$-decay threshold. Angular distributions were measured and compared to DWBA calculations to extract angular momentum transfers and $^{10}\mathrm{B}\left(3^+\right)+n$ spectroscopic factors. The recently observed and heavily discussed resonance at 11.4 MeV in $^{11}$B was not observed in this work. This result is consistent with the interpretation that it is predominantly a $^{10}\mathrm{Be}\left(0^+\right)+p$ resonance with a possible additional $^{7}\mathrm{Li}+\alpha$ contribution. The predicted $^{10}\mathrm{B}\left(3^+\right)+n$ resonance at 11.6 MeV, analogous to the 11.4-MeV proton resonance, was not observed either. Upper limits for the $^{10}\mathrm{B}\left(3^+\right)+n$ spectroscopic factors of the 11.4-MeV and 11.6-MeV states were determined. In addition, supporting configuration interaction shell model calculations with the effective WBP interaction are presented.

\end{abstract}


\maketitle

\section{Introduction}

With the possibility of studying nuclei close to the particle-emission thresholds and particle driplines at stable and rare isotope beam facilities, there have been substantial theoretical efforts to treat the nucleus as an open quantum system (see, {\it e.g.}, \cite{Oko03a, Mic10a, Baz23a, Vol24a} and references therein). Especially, the importance of including coupling to the continuum was pointed out to accurately describe near-threshold physics and when dealing with reactions with weakly bound nuclei. 

The well-bound $^{11}$B nucleus is an example where its resonance region, i.e., above the particle-emission thresholds, has been of significant interest in recent years. This interest largely originated from reports of pronounced differences in branching ratio calculations of $\beta^-$ delayed proton emission, a rare type of decay, compared to earlier measurements in $^{11}$Be \cite{Riisager14}. The large $\beta^-p$ branching ratio \cite{Riisager14}, significantly larger than expected by any previous theoretical calculation, was confirmed in \cite{Ayyad19}. Ayyad {\it et al.} explained their experimental data by the existence of a then still unobserved narrow proton resonance in $^{11}$B, located just above the proton-separation threshold ($S_p = 11228.6(5)$\,keV) with either $J^{\pi}=1/2^+$ or $3/2^+$.




Two independent experiments recently succeeded to confirm the existence of this predicted proton resonance at $E_{x} \approx 11.4$\,MeV \cite{Lopez, Ayyad2022}. One experiment used the $^{10}$Be$(d,n){}^{11}$B reaction in inverse kinematics at Florida State University's John D. Fox Superconducting Linear Accelerator Laboratory, where the radioactive $^{10}$Be beam was produced using RESOLUT \cite{Lopez}. The excitation energy was determined to be 11.44(4) MeV with a tentative $J^{\pi}=1/2^+$ ($\ell=0$) spin-parity assignment and spectroscopic factor of 0.27(6). The other experiment used proton elastic scattering with a $^{10}$Be beam produced at the ReA3 facility of the National Superconducting Cyclotron Laboratory at Michigan State University \cite{Ayyad2022}. Ayyad {\it et al.} reported an excitation energy of 11.40(2), a $1/2^+$ spin-parity assignment, and a proton partial width of $4.5\pm1.1$\,keV determined from an R-matrix analysis. The possibly weak $\gamma$-decay branch of this resonance was studied in \cite{Bot24a}. No conclusive evidence for its $\gamma$ decay was presented. An experimentally yet unresolved question is whether the newly-observed 11.4-MeV resonance also has significant $\alpha$ or neutron widths as it is located above the $\alpha$- (8664.1(1)\,keV) and close to the neutron-emission thresholds (11454.12(16)\,keV).

\begin{figure*}[t] 
  \centering
   \includegraphics[width=1.0\linewidth]{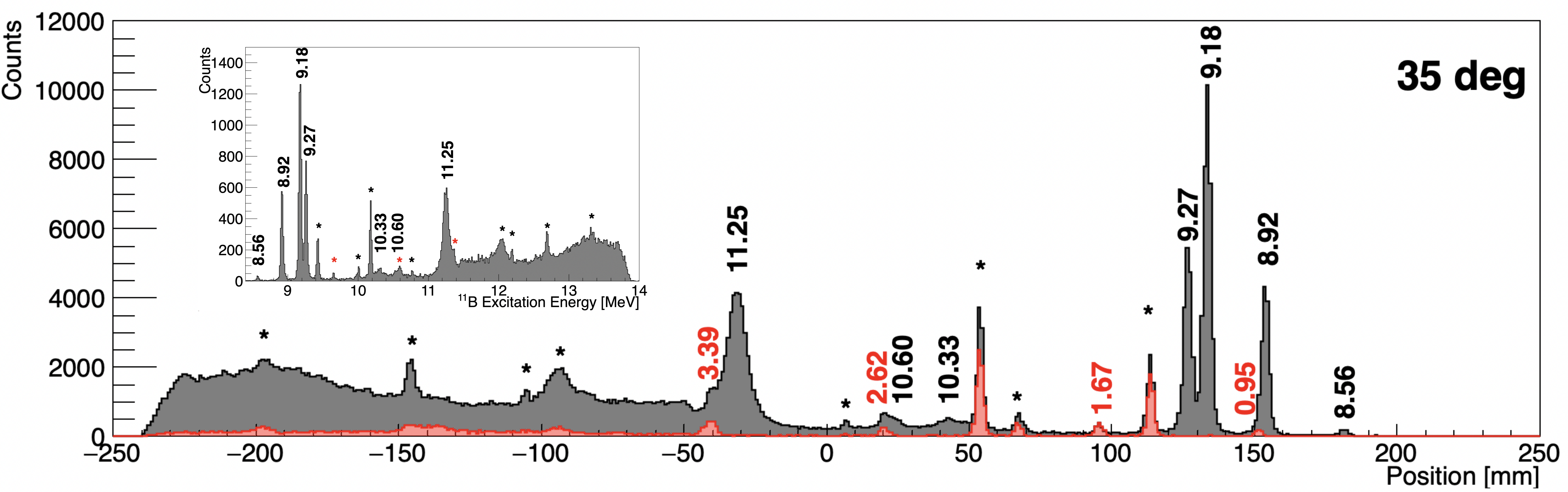} 
   \caption{SE-SPS focal plane position spectra of protons measured at $\theta_{\mathrm{SE-SPS}} = 35^{\circ}$. The gray histogram is from $(d,p)$ reactions on the target primarily composed of $^{10}$B, and the red histogram is from $(d,p)$ reactions on the target primarily composed of $^{11}$B. States labeled with black numbers, corresponding to their excitation energies in MeV, are from $^{11}$B, with red labels from $^{12}$B with their corresponding excitation energies, and states with black asterisks (*) are known contaminants from $(d,p)$ reactions on $^{12}$C and $^{16}$O. The inset shows the energy-calibrated spectrum for the $^{10}$B target, where the \nuc{12}{B} states are highlighted with red asterisks.}
   \label{fig:position}
\end{figure*}

Prior to the latest experiments, the structure of the 11.4-MeV, presumably dominant proton resonance in $^{11}$B had been explored theoretically (see, {\it e.g.}, \cite{Okolowicz20, Volya_2020, ELKAMHAWY2021, Okolowicz_2022, Nguyen2022} and references therein). For example, using shell model embedded in the continuum (SMEC) calculations, the authors of \cite{Okolowicz20} claimed that a proton resonance near the proton threshold is not surprising and that the $\beta ^-$ decay could be understood as a quasi-free decay of the $^{11}$Be halo neutron into a single proton state coupled to a $^{10}$Be core. The existence of a previously observed resonance at 11.6 MeV (see discussion in \cite{CUSSON1966481,ZWIEGLINSKI1982301,Ary85a} and references therein), conveniently located above the neutron-separation energy in $^{11}$B ($S_n = 11.4541(16)$ MeV), led the authors of \cite{Okolowicz20} to also predict a near-threshold neutron resonance similar in structure to the 11.4-MeV proton resonance. To confirm its neutron character, Okolowicz {\it et al.} called for a re-measurement of the $^{10}$B$(d,p){}^{11}$B reaction as no information for states with excitation energies in excess of $S_n$ was adopted in \cite{KELLEY201288, ENSDF} from previous $(d,p)$ experiments. They predict a $J^{\pi}=5/2^+$ neutron resonance with a narrow width of $\Gamma_n = 4$\,keV and the $l=2$ partial wave to dominate the ${}^{10}$B$+n$ decay. 

To address the experimentally open question of the dominant proton character of the 11.4-MeV resonance and to test the predictions of a possible narrow neutron resonance above $S_n$, we decided to follow the suggestion of \cite{Okolowicz20} and performed a measurement of the $^{10}$B$(d,p){}^{11}$B reaction with the Florida State University Super-Enge Split-Pole Spectrograph at the John D. Fox Superconducting Linear Accelerator Laboratory \cite{God21a, Con24a}. Specifically, we focused on the excitation energy range from the $\alpha$-decay threshold to above the neutron-separation threshold ($E_x = 8.4-13.6$\,MeV) in $^{11}$B.

\section{Experimental Details}

The experiment was performed at Florida State University’s John D. Fox Superconducting Linear Accelerator Laboratory \cite{fox_lab}. A 16-MeV deuteron beam was accelerated by the 9-MV Pelletron-charged Super-FN-Tandem Van de Graaff accelerator. The deuteron beam impinged on two different boron targets. The first was a self-supporting $^{10}$B target containing $^{11}$B, $^{12}$C, and $^{16}$O contaminants with a total density of about 70 $\mu$g/cm$^2$. The \nuc{11}{B} contamination was on the order of $8\%$. To assist in the identification of states coming from the $^{11}$B$(d,p){}^{12}$B reaction caused by the $^{11}$B target contamination, data were taken at each angle with a second target enriched in $^{11}$B ($>99\%$) with a density of about 70 $\mu$g/cm$^2$ on a 20 $\mu$g/cm$^2$ carbon backing. The targets were installed in the sliding-seal scattering chamber upstream of the Super-Enge Split-Pole Spectrograph (SE-SPS) \cite{God21a, Con24a}, where its two dipole magnets focused the charged particles toward the focal-plane detector. The detector includes a position-sensitive ionization chamber filled with isobutane gas, and a plastic scintillator for measuring the remaining energy of the particles after passing through the ionization chamber. Particle trajectories can be reconstructed and momentum spectra can be constructed by using the acquired position information from position-sensitive anode pick-up pads with delay-line readout. Particle (reaction) identification is achieved by measuring the energy loss in the ionization chamber and the remaining energy with the plastic scintillator. The solid-angle acceptance of the SE-SPS was set at 4.6 msr for this experiment. To determine differential cross sections at different scattering angles in the laboratory frame, measurements were performed from $10^{\circ}$ to $50^{\circ}$ in $5^{\circ}$ steps.

\begin{figure*}[t] 
   \centering
   \includegraphics[width=.8\linewidth]{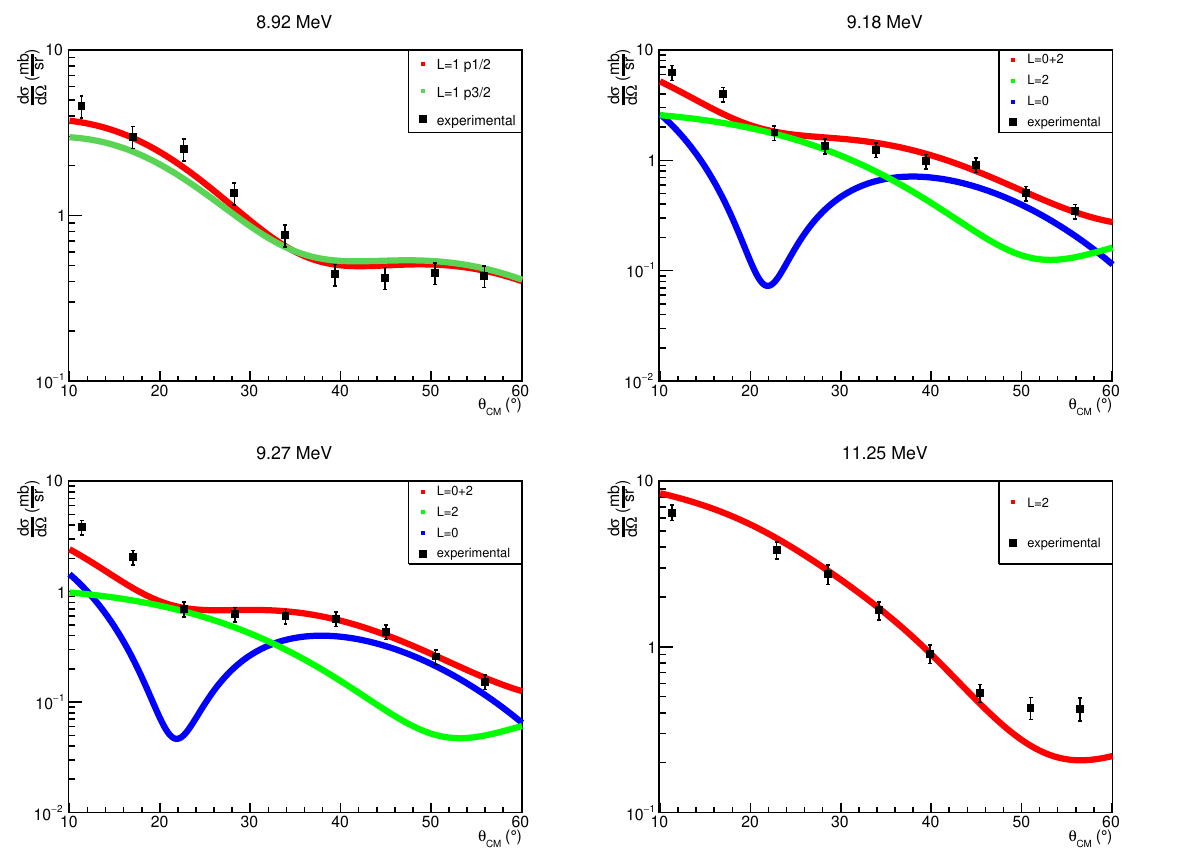} 
   \caption{Measured $(d,p)$ angular distributions for states populated in $^{10}$B$(d,p){}^{11}$B are shown (symbols with error bars). DWBA calculations, scaled according to the determined spectroscopic factors, are overlaid on the data. Some details of the DWBA calculations are given in the text.}
  \label{fig:AngDis}
\end{figure*}

The magnetic field of the spectrometer was set to measure the $^{11}$B excitation energy range from about 8.4 MeV to 13.6 MeV. An example of a focal plane position spectrum, gated on the $(d,p)$ reaction and measured at $\theta_{SE-SPS} = 35^{\circ}$, is shown in Fig. \ref{fig:position}. The gray $(d,p)$ spectrum was measured with the enriched $^{10}$B target, while the red spectrum was measured for the enriched $^{11}$B target to identify $^{12}$B contaminants in the former. The lowest excited state of $^{11}$B we observe with the chosen magnetic setting is at $E_x$ = 8.56 MeV. At least six excited states of $^{11}$B above its $\alpha$-decay threshold of 8664.1(1)\,keV were observed with four of them isolated well enough to study angular distributions (see Fig.\,\ref{fig:AngDis}) and extract spectroscopic factors (see Table\,\ref{table:Results}).

Differential cross sections were calculated as a function of the center-of-mass (CM) angle and compared to Distorted Wave Born Approximation (DWBA) calculations performed with the \textsc{fresco} code \cite{fresco} (see Fig.\,\ref{fig:AngDis}). For the DWBA calculations, optical model parameters for the deuteron channel were taken from \cite{Daehnick} and from \cite{Koning} for the proton channel. A $\chi^2$ minimization was performed to determine the spectroscopic factors given in Table\,\ref{table:Results}.

\section{Discussion}

We observe a triplet of $^{11}$B states around 9\,MeV. In agreement with its adopted $J^{\pi} =5/2^-$ spin-parity assignment, we observe the 8.92-MeV to be populated through an $\ell=1$ angular momentum transfer. This is also in agreement with the angular momentum transfer reported in \cite{HINDS1962114, POORE196797} for previous $(d,p)$ experiments. The $\ell=1$ angular momentum transfer indicates that the neutron is either transferred into the $0p_{1/2}$ or $0p_{3/2}$ orbital. Both possibilities are shown in Fig. \ref{fig:AngDis}. A better fit is obtained with the neutron transferred into $0p_{1/2}$ orbital. The conflicting $(d,p)$ angular momentum transfer of a mixed $\ell = 0+2$ transfer, reported in \cite{Bilaniuk}, should be dropped.  

The angular distributions measured for the 9.18-MeV and 9.27-MeV states were, however, both best described by a combination of $\ell=0$ and $\ell=2$ angular momentum transfers. The individual contributions, determined through $\chi^2$ minimization, as well as their sum are shown in Fig. \ref{fig:AngDis}. Previous work \cite{Bilaniuk, HINDS1962114, POORE196797} reported that these states were populated through pure $\ell=0$ transfers. Our new data clearly show that an $\ell =0$ transfer is not sufficient to describe the experimental data. The addition of an $\ell=2$ component improves the agreement with data significantly. When inspecting the angular distributions presented in \cite{Bilaniuk, HINDS1962114, POORE196797}, it can be clearly seen that the $\ell=0$ calculations alone also undershoot the shallow minima in the experimental angular distributions for both states. This was ignored in previous publications and should be corrected in the adopted data \cite{KELLEY201288, ENSDF}. The spectroscopic factors for the individual contributions are listed in Table\,\ref{table:Results}.

\begingroup
\begin{table}[!t]
\caption{\label{table:Results} Excitation energies, adopted spin-parity assignments \cite{KELLEY201288, ENSDF}, measured orbital angular momentum transfers, and spectroscopic factors determined from comparison to DWBA calculations for states populated in $^{10}$B$(d,p){}^{11}$B. The corresponding $(d,p)$ angular distributions are shown in Fig.\,\ref{fig:AngDis}.}
\begin{ruledtabular}
\begin{tabular}{cccc}
$E_{x}$ [MeV] & $J^{\pi}$ & $\ell$& $S$ \\
\hline
8.92& $5/2^-$ & 1 & 0.13(2) \\
9.18 & $7/2^+$ & 0+2 & 0.08(1)+0.18(3) \\
9.27 & $5/2^+$ &  0+2 & 0.06(1)+0.10(2) \\
11.25(1)\footnotemark &$(9/2)^+$\footnotemark & 2 &  0.24(4)
\footnotetext[1]{This energy was determined from a linear position-to-energy calibration using the other known excited states of $^{11}$B seen in Fig.\,\ref{fig:position}.}
\footnotetext[1]{This is possibly the adopted 11.27-MeV, $J^{\pi}=9/2^+$ state \cite{KELLEY201288,ENSDF}. See text for further discussion.}
\end{tabular}
\end{ruledtabular}
\end{table}
\endgroup

Two states between $E_{x}=10-11$ MeV appear to be weakly populated in the $^{10}$B$(d,p){}^{11}$B reaction (see Fig.\,\ref{fig:position}). The adopted excited state at 10.330(8) MeV, which has a significant width\,\cite{KELLEY201288, ENSDF}, has presumably been populated in an earlier $(d,p)$ experiment \cite{Elkind}. Unfortunately, at many scattering angles, the protons, which would indicate the unambiguous population of this state, are located in a position largely overlapping with the 3.85-MeV contaminant state from the $^{12}$C$(d,p){}^{13}$C reaction. At larger angles, when the state has more separation from the $^{13}$C state, a broad structure shows up which could be the adopted 10.33-MeV state or a possible combination of it with the adopted 10.26-MeV state. Both have sizable widths. Similarly, we see evidence of a state being populated at $E_x = 10.6$\,MeV. However, this state partly overlaps with the 2.62-MeV $^{12}$B state in the focal plane (see Fig.\,\ref{fig:position}). While these states, thus, appear to be populated in the $(d,p)$ reaction, it is difficult to extract reliable spectroscopic information for them.

A strong state just above the proton-emission threshold of $S_p = 11.2285(5)$\,MeV \cite{ENSDF} was observed in this work (see Fig.\,\ref{fig:position}). It has not been reported in any previously published $(d,p)$ work. The three low-lying states around 9\,MeV and other contaminant states were used as calibration points in a linear fit to determine the energy of states with higher energy. This state was determined to have an excitation energy of 11.25(1)\,MeV and total width of $108(12)$\,keV, typical for many adopted states in this energy range \cite{KELLEY201288, ENSDF}. The $(d,p)$ angular distribution is well described by an $\ell=2$ angular momentum transfer (see Fig. \ref{fig:AngDis}). If the neutron is assumed to be transferred into the $0d_{5/2}$ orbital, a spectroscopic factor of 0.24(4) is determined (see Table\,\ref{table:Results}). This is likely the adopted 11.272-keV state, as both the measured angular momentum transfer and width would agree with the adopted spin-parity assignment of $J^{\pi} = 9/2^+$ and adopted width of $110(20)$ keV. However, based on our data alone, a large spin range of $J^{\pi} = 1/2^+ - 11/2^+$ is possible. There is a previous mention of the 11.2-MeV state's population in $(d,p)$ in an abstract \cite{Groce}. This abstract indicates its observation as well as a width of $<30$ keV. No further information was provided though and, to our knowledge, these results were never published. In principle, the work of Ref. \cite{Elkind} covered the excitation energy range to see this state in the $(d,p)$ reaction. But Elkind {\it et al.} only performed measurements at $90^{\circ}$. At this angle, the peak would possibly overlap with the 4.56-MeV $^{17}$O state populated in $^{16}$O$(d,p){}^{17}$O, a significant contaminant which they reported \cite{Elkind}. We can exclude a prominent contribution of a possible $^{17}$O contaminant to the 11.25-MeV state observed in our work.



\begin{figure}[t] 
   \includegraphics[width=1.0\linewidth]{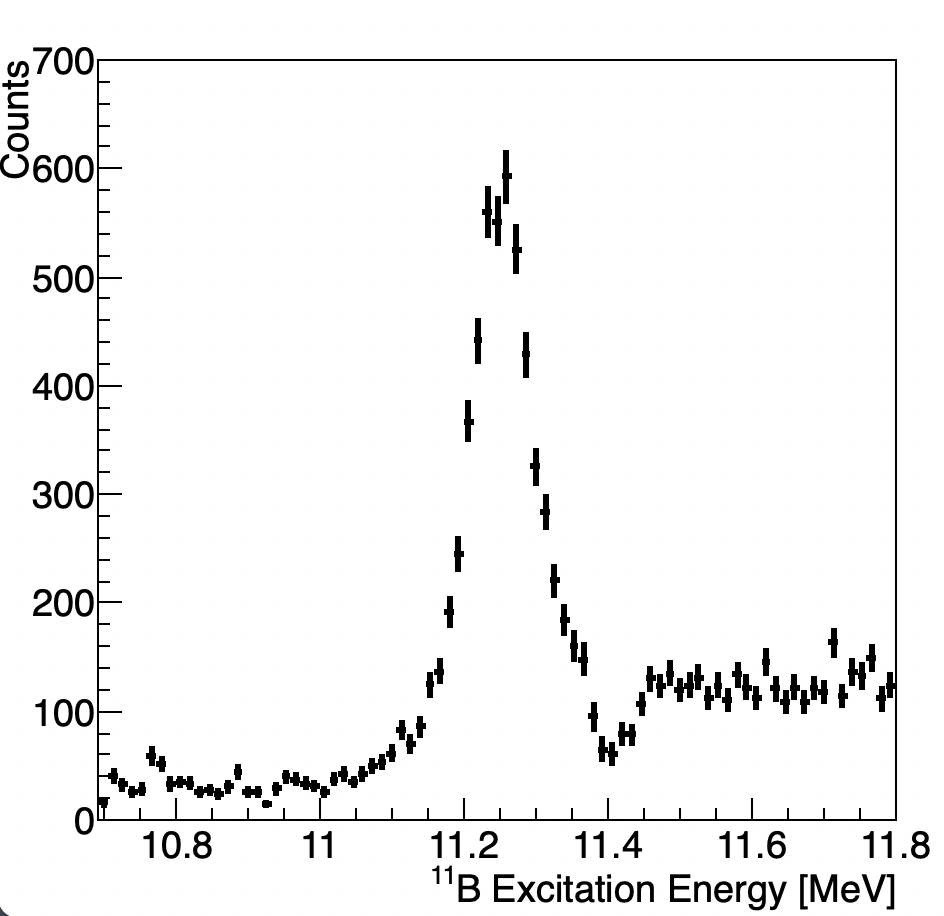} 
   \caption{$^{11}$B excitation energy spectrum between $E_x = 10.7$\,MeV and 11.8\,MeV measured at $\theta_{\mathrm{SE-SPS}} = 35^{\circ}$. The spectrum was obtained by subtracting the $(d,p)$ data measured with the enriched $^{11}$B target from the data measured with the enriched $^{10}$B target (see also Fig.\,\ref{fig:position}). Details about the subtraction and normalization procedure are provided in the text.}
   \label{fig:subtraction}
\end{figure}

In the high-energy tail of the 11.25-MeV state, a peak-like structure was observed (see Fig.\,\ref{fig:position}). From our complementary measurements with the $^{11}$B target, we determined that this peak corresponds to the known 3.389-MeV excited state of $^{12}$B. As this contaminant peak lies close to where the much discussed 11.4-MeV resonance \cite{Ayyad19, Ayyad2022, Bot24a} would be expected in the focal plane, if it had any $^{10}\mathrm{B}+n$ component, we removed the $^{12}$B contaminant through subtraction. To do so, the two datasets were normalized to the yield of the clearly resolved 1.67-MeV state in $^{12}$B, respectively (see Fig.\,\ref{fig:position}). The resulting spectrum after subtraction is shown in Fig. \ref{fig:subtraction}. It is clear that the 11.4-MeV state is not populated strongly; if at all. To get an upper limit on the possible population of a $^{11}$B state at 11.4 MeV, the number of counts above the background was collected at each angle and a spectroscopic factor determined in the same way as for the other states. We assumed that the state could only be populated through an $\ell = 2$ angular momentum transfer because of its proposed $J^{\pi} = 1/2^+, 3/2^+$ assignment. An upper limit of 0.02 was obtained for the spectroscopic factor. This is consistent with interpreting it as a strong proton resonance with possible additional $^{7}\mathrm{Li}+\alpha$ strength. Based on our data, we can, thus, conclude that there is no significant $^{10}\mathrm{B}+n$ component for the 11.4-MeV state.

The 11.6-MeV state \cite{KELLEY201288, ENSDF}, which was considered to be the analogous neutron resonance to the 11.4-MeV proton resonance due to its location relative to the neutron threshold $S_n = 11454.12(16)$\,keV in $^{11}$B by Okolowicz {\it et al.} \cite{Okolowicz20}, is not strongly populated in our work either. In fact, there is no indication of its population at all. As for the 11.4-MeV state, an upper limit of 0.013 for its spectroscopic factor was determined assuming $\ell = 0$ and $\ell =2$ angular momentum transfers separately and scaling the predicted distributions to the differential cross sections at the most forward angles.

Therefore, based on our new data, we conclude that there is no narrow state close to $S_n$ in $^{11}$B which appears to be strongly populated in $(d,p)$. This clearly challenges the predictions of Okolowicz {\it et al.} that such a narrow neutron resonance should possibly be observed in one-neutron transfer experiments. If any other states were populated at higher excitation energy, they would have to be broad and/or weakly populated. Our measured spectra (see Fig.\,\ref{fig:position}) might indeed support that many broad resonances contribute to the observed ``background'' at higher excitation energies. In general, the level density and width of states seem to increase significantly past $S_p$ \cite{ENSDF}. An experiment with an ultra pure target or, possibly, in inverse kinematics with clean beam selection would be able to test this possibility in more detail as it would not suffer from target contamination.
\\


To gain some additional insight and possibly guide the planning of future experiments, we performed configuration-interaction shell-model calculations using the effective WBP Hamiltonian of \cite{Warburton} in a model space, which allowed one-particle-one-hole (1p-1h) excitations from $0s$ to $0p$ and from $0p$ to $1s0d$ orbitals, respectively. 2p-2h and $\alpha$-cluster configurations are not considered. We focus on positive-parity states at higher excitation energies and examine their $\ell= 0$ and $\ell= 2$ spectroscopic factors for possible $^{10}\mathrm{B}\left(3^+\right)+n$ and $^{10}\mathrm{Be}\left(0^+\right)+p$ contributions to their wavefunctions.



\begingroup
\begin{table}[!t]
\caption{\label{table:SM} Shell model predictions for $^{10}\mathrm{B}\left(3^+\right)+n$ and $^{10}\mathrm{Be}\left(0^+\right)+p$ spectroscopic factors of positive-parity states in $^{11}$B with $E_x \leq 13$\,MeV. Spectroscopic factors are given for the angular momentum transfers, which would populate the states from the $^{10}$B and $^{10}$Be ground states in one-nucleon transfer reactions, respectively. See text for further discussion.}
\begin{ruledtabular}
\begin{tabular}{cccccc}
$J^{\pi}$ & $E_{x, th.}$ & \multicolumn{2}{c}{$S\left(^{10}\mathrm{B}\left(3^+\right)+n\right)$} & \multicolumn{2}{c}{$S\left(^{10}\mathrm{Be}\left(0^+\right)+p\right)$}  \\
 & [MeV] & $\ell = 0$ & $\ell = 2$ & $\ell = 0$ & $\ell = 2$ \\
 \hline
$1/2^+_1$  & 6.36  &   -   & 0.01 & 0.27    & -  \\
$1/2^+_2$  & 10.87 &   -   & 0.02 & 0.05        & -   \\
$1/2^+_3$  & 12.53 &   -   & 0.00 & 0.05         & -  \\
$1/2^+_4$  & 12.89 \footnotemark &   -   & 0.09 & 0.16   & -   \\
$1/2^+_5$  & 12.92 &   -   & 0.00 &   0.25      & -   \\
\hline
$3/2^+_1$  & 8.05  &   -   & 0.00 & -    & 0.10   \\
$3/2^+_2$  & 9.73  &   -   & 0.00 & -    & 0.03   \\
$3/2^+_3$  & 10.84 &   -   & 0.00 & -    & 0.01   \\
$3/2^+_4$  & 11.16 &   -   & 0.02 & -    & 0.03   \\
$3/2^+_5$  & 11.60 &   -   & 0.02 & -    & 0.00   \\
$3/2^+_6$  & 12.13 &   -   & 0.00 & -    & 0.00  \\
$3/2^+_7$  & 12.88 &   -   & 0.01 & -    & 0.04  \\
\hline
$5/2^+_1$  & 6.71  & 0.02 & 0.03 & - & 0.26   \\
$5/2^+_2$  & 8.34  & 0.40 & 0.22 & - & 0.02  \\
$5/2^+_3$  & 9.87  & 0.03 & 0.00 & - & 0.01 \\
$5/2^+_4$  & 11.22 & 0.04 & 0.20 & - & 0.03 \\
$5/2^+_5$  & 11.51 & 0.04 & 0.23 & - & 0.01 \\
$5/2^+_6$  & 12.70 & 0.00 & 0.00 & - & 0.09  \\
\hline
$7/2^+_1$  & 8.70  & 0.09 & 0.46 & -  & -  \\
$7/2^+_2$  & 10.05 & 0.09 & 0.03 & -  & -  \\
$7/2^+_3$  & 11.22 & 0.43 & 0.00 & -  & -  \\
$7/2^+_4$  & 11.64 & 0.05 & 0.13 & -  & -  \\
$7/2^+_5$  & 12.79 & 0.15 & 0.11 & -  & -  \\
\hline
$9/2^+_1$  & 9.57  &   -   & 0.25 & -  & -  \\
$9/2^+_2$  & 11.52 &   -   & 0.41 & -  & -   \\
\hline
$11/2^+_1$ & 12.67 &   -   & 0.69 & -  & - 
\footnotetext[1]{This is the lowest $T=3/2$ state.}
\end{tabular}
\end{ruledtabular}
\end{table}
\endgroup

The shell model predicts numerous $1/2^+$ and $3/2^+$ states but with very little to no spectroscopic strengths. In agreement with the predictions, none of these states were seen in our $(d,p)$ experiment. Some of these states have larger proton spectroscopic factors (see Table\,\ref{table:SM}). Based on the tentative $J^{\pi}=1/2^+$ assignment and reported proton spectroscopic factor of 0.27(6) \cite{Lopez}, the 11.4-MeV resonance might correspond to the predicted 12.92-MeV state in our shell-model calculations. In general, one should consider the experimentally observed $1/2^+$ state at 11.4 MeV, however, as a linear combination of the four predicted states above 7 MeV, in which case it is possible to get a $^{10}\mathrm{Be}\left(0^+\right)+p$ spectroscopic factor as large as 0.27(6). Okolowicz {\it et al.} \cite{Okolowicz20} predicted that the $1/2^+_3$ eigenstate of their SMEC calculations corresponded to the 11.4-MeV resonance. The vanishing spectroscopic factors for $^{10}\mathrm{B}\left(3^+\right)+n$ contributions to all predicted $1/2^+$ states agree with our upper limit of 0.013, i.e., the non-observation of the 11.4-MeV state in $(d,p)$. For the excited $3/2^+$ states up to 13\,MeV of excitation energy, none of them seems to have a pronounced neutron or proton 1p-1h character.

The first $5/2^+$ state is predicted at 6.71\,MeV, about 600\,keV below the first adopted $5/2^+$ state at 7.286\,MeV\,\cite{KELLEY201288, ENSDF}. It was not within our acceptance and only small neutron spectroscopic factors are predicted. In agreement with the calculations, the 9.27-MeV, $5/2^+_2$ state and 9.18-MeV, $7/2^+_1$ state are most strongly populated in $(d,p)$. They are predicted about 900\,keV and 500\,keV below the experimentally adopted states \cite{KELLEY201288, ENSDF}. As observed in experiment, the shell model expects the states to have both $\ell = 0$ and $\ell = 2$ components in their wavefunctions. The absolute values of the predicted spectroscopic factors do not agree with the experimentally determined ones and there are no reliable values reported in literature we could compare to. It should be noted that if the fit is restricted to just the smaller angles, the $\ell=0$ component increases. For the 9.19-MeV state, the $\ell=0$ spectroscopic factor increases from 0.08 to 0.12. For the 9.28-MeV state, the $\ell=0$ spectroscopic factor increases from 0.06 to 0.12. These values are somewhat in agreement with the values determined from DWBA calculations presented in \cite{POORE196797}. In our work, we chose to perform the $\chi^2$ minimization for all measured scattering angles though. Proton spectroscopic factors are predicted to be small for these two states. We do not observe candidates for the predicted $5/2^+$ and $7/2^+$ states with $E_x > 11$\,MeV and significant neutron spectroscopic factors. However, states with this spin-parity assignment are adopted at these excitation energies. As discussed earlier, they might be too broad to result in clear peaks for us to resolve or their spectroscopic factors might be smaller than predicted. Considering that also the SMEC calculations expected narrow resonances at these energies, it might be instructive to perform calculations including proton and neutron 1p-1h, 2p-2h, as well as $\alpha$-particle-type excitations to study widths of states and their spectroscopic factors on the same footing.

In closing our discussion, we want to mention that the shell-model calculations with the WBP Hamiltonian predict the first $9/2^+$ state at 9.57\,MeV. There are indications that this state should correspond to the 11.2-MeV state, also discussed in this work. While the spectroscopic factors agree, the difference in excitation energy is quite large, which bears the question where the $9/2^+_2$ state with even larger spectroscopic strength should be expected to be observed experimentally. Currently, no candidate is known. An experimental candidate for the $11/2^+_1$ state is adopted at about 14\,MeV and expected to be a 500-keV wide resonance \cite{KELLEY201288, ENSDF}.

\section{Conclusion}

A $^{10}$B$(d,p){}^{11}$B experiment with a 16-MeV deuteron beam was performed at the FSU SE-SPS with the primary goal to test any possible $^{10}\mathrm{B}\left(3^+\right)+n$ contribution to the presumably strong $^{10}\mathrm{Be}\left(0^+\right)+p$ resonance at 11.4\,MeV. Based on the new $(d,p)$ data, a significant contribution can now be excluded as presented in this work. To test the predictions of Okolowicz {\it et al.} \cite{Okolowicz20} that, similar to this proton resonance, a narrow $^{10}\mathrm{B}\left(3^+\right)+n$ resonance should exist above the neutron-separation threshold, the excitation spectrum of $^{11}$B was measured up to an energy of about 13.6\,MeV, therefore, covering a large energy window above the proton- and neutron-separation thresholds. No evidence for the existence of such a narrow neutron resonance could be found. If it existed, it would need to be broad or the strength largely fragmented.

Up to an excitation energy of about 13.6\,MeV, at least six excited states of $^{11}$B were identified. Four of them were sufficiently resolved to extract spectroscopic information. $^{10}\mathrm{B}\left(3^+\right)+n$ spectroscopic factors and the orbital angular momentum transfer were determined from a comparison between the experimental data and DWBA calculations. $(d,p)$ data for an excited state at $E_{x}=11.25(1)$ MeV, the only clearly resolved excited state above $S_p$ and with a total width of $108(12)$ keV, are presented for the first time. The 11.25-MeV state has a clear $\ell=2$ $(d,p)$ angular distribution with a $^{10}\mathrm{B}\left(3^+\right)+n$ spectroscopic factor of 0.24(2), when assuming the neutron to be transferred into the $0d_{5/2}$ orbital. It might, thus, correspond to the adopted 11.27-MeV state with $J^{\pi} = 9/2^+$ and a total width of 110(20)\,keV \cite{KELLEY201288, ENSDF}. However, based on our data alone, a large spin range of $J^{\pi} = 1/2^+ - 11/2^+$ is possible.

After addressing the possible $^{10}\mathrm{B}\left(3^+\right)+n$ contribution to the 11.4-MeV resonance in this work, outcomes of experiments already performed to study the predicted $^{7}\mathrm{Li}+\alpha$ contribution are awaited. A possibility to further study the weak $\gamma$-decay branch of this resonance \cite{Bot24a} could be to couple the SE-SPS with an array of highly-efficient CeBr$_3$ detectors \cite{Con24a} in, {\it e.g.}, the $^{7}\mathrm{Li}(^{7}\mathrm{Li},t)^{11}\mathrm{B}$ reaction. Calculations addressing the non-observation of the predicted 11.6-MeV narrow neutron resonance are called for.


\begin{acknowledgments}
The authors wish to thank Matt Gott for assistance with the targets which were provided by the Center for Accelerator Target Science at Argonne National Laboratory and Nicholas Keeley for assistance with DWBA calculations. This work was supported by U.S. National Science Foundation (NSF) under Grants Nos. PHY-2012522 [WoU-MMA:
Studies of Nuclear Structure and Nuclear Astrophysics (FSU)] and PHY-2110365 [Nuclear Structure Theory and its Applications to Nuclear Properties, Astrophysics and Fundamental Physics (MSU)].
\end{acknowledgments}

\bibliography{11B_bib}

\end{document}